\documentclass[conference]{IEEEtran}
\IEEEoverridecommandlockouts

\usepackage{cite}
\usepackage{spverbatim}
\usepackage{subfig}
\usepackage{amsmath,amssymb,amsfonts}
\usepackage{algorithmic}
\usepackage{graphicx}
\usepackage{textcomp}
\usepackage{xcolor}
\usepackage{fancyhdr}
\def\BibTeX{{\rm B\kern-.05em{\sc i\kern-.025em b}\kern-.08em
    T\kern-.1667em\lower.7ex\hbox{E}\kern-.125emX}}

\begin{document}

\title{DIMSIM -- \textbf{D}evice \textbf{I}ntegrity \textbf{M}onitoring through \textbf{iSIM} Applets and Distributed Ledger Technology}

\author{\IEEEauthorblockN{ Tooba Faisal \& Emmanuel Marilly}
\IEEEauthorblockA{\textit{Nokia Bell-Labs, France }
\\
\textit{Tooba.Faisal@nokia-bell-labs.com}
\\
\textit{Emmanuel.Marilly@nokia-bell-labs.com}
}}
\maketitle
\thispagestyle{fancy}
\begin{abstract}
 
In the context of industrial environment, devices, such as robots and drones, are vulnerable to malicious activities such device tampering (e.g., hardware and software changes). The problem becomes even  worse in a multi-stakeholder environment where multiple players contribute to an ecosystem.

 In such scenarios, particularly, when devices are deployed in remote settings, ensuring device integrity so that all stakeholders can trust them is challenging. Existing methods, often depend on additional hardware like the Trusted Platform Module (TPM) which may not be universally provided by all vendors. In this study, we introduce a distributed ledger technology-oriented architecture to monitor the remote devices' integrity using eUICC technology, a feature commonly found in industrial devices for cellular connectivity. We propose that using  secure applets in eUICC, devices' integrity can be monitored and managed without installing any additional hardware.
 
 To this end, we present an end-to-end architecture to monitor device integrity thereby enabling all the stakeholders in the system to trust the devices. Additionally, we leverage the  properties of immutable databases to provide robustness and efficiently to our model. In our primary evaluations, we measure the overhead caused by hashing our proposed data packets and performance of integrating an immutable database into our system. Our results show that performing hashing on our data packets takes order of microseconds, while reading and writing to an immutable database also requires only milliseconds.

\end{abstract}

\section{Introduction}

Future industrial systems, such as Industry 4.0 and Industry 5.0, are envisioned as multi-stakeholder environments— a multi-actor and open ecosystem~\cite{redlich2015strategy} — in which assets and services may not necessarily come from a limited number of providers. These systems provide  opportunities for a wide range of vendors, regardless of their business size, to contribute to the ecosystem. The primary concern in such systems is trust among the stakeholders within the ecosystem~\cite{accountable_OCN, faisal2023accountable}.
 
We believe that such open systems can only be enabled \textit{if and only if} the systems are inherently \textit{trustable} and \textit{accountable}.

This assertion is grounded in the following primary reasons:
The first challenge related to open systems~\cite{redlich2015strategy} is their ability to easily interoperate with devices from a wide range of providers and form end-to-end heterogeneous systems. In such heterogeneous systems various device providers will be involved in an end-to-end service. Therefore, each party depends on other parties for the smooth operations and quality of service delivered. Hence, malfunctioning of a single device will impact the performance of an overall service. The challenge is not limited to quality but also for security reasons; a corrupt or poorly secured device can introduce dangers to the full system~\cite{botnets}. This necessitates robust monitoring and reporting mechanisms that allow complete transparency among actors, thereby establishing trust.

The second challenge revolves around establishing accountability for devices in large-scale deployments featuring multiple participants in the ecosystem. In industrial contexts, ensuring the accountability and reliability of remote devices poses a significant challenge. When devices operate remotely, their performance may deviate from the intended programming. This discrepancy could stem from various factors, including malicious interference by external entities altering device software to compromise the environment or derive personal gains. Thus, devices integrated into industrial environments must inherently embed mechanisms ensuring not only the integrity of their software and firmware but also providing unequivocal assurances regarding the accuracy and accountability of the data they generate. A robust system of accountability is crucial to uphold the reliability and trustworthiness of the entire ecosystem.

Hence, future generation of industrial systems must be both open and accountable, inherently ensuring trust. Data generated by these devices  and transactions between stakeholders must be accurate, trustworthy, and reliable for all the participants.  Additionally, these systems should guarantee the proper execution of devices' programmed functions, while also detecting and taking appropriate actions for any malicious behaviors.  The entire ecosystem should operate autonomously, functioning in a zero-touch manner. 

Current systems rely on Trusted hardware such as trusted platform modules, with remote attestation~\cite{Nokia_AE}. Such architectures can both perform local and remote confirmation of the device integrity. However, the problem with such architectures is that an additional hardware such as Trusted Platform Module must be installed inside a device. Moreover, in the case of remote attestation, when a party verifies the device integrity, the system relies on a centralized architecture of remote attestation. This means, when multiple parties want to keep track of device integrity, they must rely on a central party. 

To this end, our work introduces an end-to-end and distributed device integrity monitoring system designed for a multivendor environment. The system monitors device integrity and enables trust, without the need for additional hardware: DIMSIM (Device Integrity Monitoring with SIM Applets)~(Fig.\ref{fig:DIMSIM}). We leverage eUICC technology to maintain device integrity. eUICC is not an additional burden on the device and is a standard to provide connectivity to remote devices~\cite{gsma_euicc_2}. The remote verification entity in our proposal is collectively managed by all participants within the ecosystem and utilizes immutable records management. Recent application of immutable database \emph{immudb} is a new class of databases that is auditable, irrepudiable and temper-resistant by design~\cite{paik2020immudb}. This ensures that all stakeholders can access the records, and tampering with them is not possible.
On the other hand, an eUICC, or embedded Universal Integrated Circuit Card, is closely related to the eSIM (embedded Subscriber Identity Module) or iSIM (integrated Subscriber Identity Module).

An eUICC is a component that can host multiple SIM profiles, allowing for the dynamic switching and reconfiguration of mobile network providers, profiles or high security services / elements additionally to the telecom services~\cite{gsma_euicc}. We propose upholding device integrity through the utilization of these secure elements, unveiling the Attestation Applet, a novel secure element nestled within the eUICC.

DIMSIM is a modular system leveraging the combination of the eUICC (eSIM or iSIM) technology at the device level, and Distributed Ledger technology at the cloud level (Fig.~\ref{fig:DIMSIM}). We exploit eUICC in combination with distributed ledger technology to provide all the stakeholders a transparent view of the device software and firmware. We use a specific type of distributed ledgers --  ``Permissioned Distributed Ledger (PDL)'' due to that the fact they have already been discussed widely in industrial applications, for example, in operations and control networks~\cite{accountable_OCN}, automated monitoring and network devices' sharing~\cite{faisal2023accountable}.


\section{Related Work}

In this work, we introduce a system designed to facilitate a multivendor environment. Our objective is to enable stakeholders to precisely assess the integrity of each individual device. We achieve this with the combination of eUICC and  distributed ledger technologies. 

Several systems to monitor device integrity are proposed such as $I^3$~\cite{patil2004i3fs} and Tripwire~\cite{kim1994design}. Such proposals enable administrators to verify device integrity but do not address the concern of malicious device deliberately tampering with the device.

An alternate to Trusted Platform Module (TPM) within the domain of eUICC (embedded Universal Integrated Circuit Card) is proposed by Chakraborty et al.~\cite{chakraborty2019simtpm} in \textit{SimTPM}. They proposed incorporating TPM functionality in eUICC. SimTPM utilizes the eUICC's non-volatile memory to store device measurements. Unlike us, they use  TPM commands for integrity monitoring and do not perform real-time monitoring. Neither simTPM automatically reports device changes to a remote location (e.g., device lessee).

In the context of mobile devices, Raj et al. introduced \textit{firmwareTPM (fTPM)}~\cite{raj2016ftpm}, a solution specifically tailored for mobile devices that relies on ARM TrustZone for its operation.

Petroni et al.'s \textit{Copilot}~\cite{petroni2004copilot} aligns closely with our proposal, although it necessitates the use of a PCI add-in card. This PCI card plays a critical role in detecting malicious modifications to the kernel.

Our solution surpasses the limitations of TPM functionalities and eliminates the dependency on additional hardware. Notably, it's important to mention that eUICC is already a standard component in devices requiring connectivity. Our framework, DIMSIM, represents an end-to-end solution that continuously monitors device integrity and reports any anomalies detected to the stakeholders.

Furthermore, DIMSIM offers a transparent view to all stakeholders and ability to stop the device, if an anomaly is detected. Such functionalities are crucial for ensuring the viability of future industrial applications. 
\section{Threat Model}
The threat model focused on a heterogeneous environment where customers and solution providers procure assets, such as drones and cloud services from a diverse set of vendors. To enable trust and transparency in such a multistakeholder environment, accurate and trustable monitoring of asset usage, service delivery and device health is required, due to SLA assurance such as agreed upon service quality and accurate charging. Assets such as drones can be monitored through measurement software installed within devices. This monitoring software can be stored in a shared repository and available for audit by stakeholders. In our threat model, all parties have potential motives for malicious behavior. Device vendors may attempt to inflate usage readings to unfairly charge solution providers, while customers may underreport usage to minimize their bills. The actors can tamper with devices and other data storage locations (e.g., edge) by manipulating with the measurement software and/or measurements data for charging purposes. Despite auditable software, hidden files and external software can enable malicious device behavior.

\section{Architecture}

Our system proposal is to provide device integrity monitoring without any additional hardware. Our proposal has a distributed architecture, in which the device integrity is monitored by a secure applet. The remote location verification is performed only in the situations of dispute. 
Three different types of entities interact with this system a) \textbf{Solution Provider} -- an entity which forms solutions by combining different services (e.g., connectivity) and devices (e.g., robots), b) \textbf{Device Vendor} -- device provider who leases their devices to solution providers, and  c) \textbf{Service Provider} - entities which provides services such as connectivity and robot controller.

In addition to the actors, there are following  main components for our architecture:
\begin{itemize}
    \item \textbf{Assets} -- For example, devices and services available on a platform such as a marketplace. The devices have cellular connectivity, for example, private 5G. 
    
    \item \textbf{Attestation Applet} -- our proposed novel secure element which monitors the device integrity. The Applet is controlled by the solution provider. The attestation applet ensures device integrity by continuous monitoring, reporting anomalies to the remote verifier. 
     
    \item \textbf{Remote Verifier} -- an immutable and untemperable entity managed by all the stakeholders  It maintains and manages immutable database and record of all good-known values of devices software and firmware. However, the Remote Verifier cannot update the records themselves. They must take consensus from all the stakeholders to update the records. The remote verifier may be running on a platform such as edge cloud.
    
    \item \textbf{Permissioned Distributed Ledger (PDL)} -- a distributed ledger managed by all the stakeholders within the ecosystem. We use a permissioned type of distributed ledger, which is managed by the consortium of stakeholders within the ecosystem. The transactions and contracts among the participants of the ecosystem are recorded in the PDL, such as service level agreements. The ledger is consensus agnostic. 
\end{itemize}

Our system is formed by participants running a permissioned distributed ledger (PDL) that includes agreements between the actors within the ecosystem. Assets (e.g., services, devices)  and their associated contracts (e.g., service level agreements~(SLAs)) are deployed as smart contracts on the PDL. When a solution provider forms a solution by leasing devices, service level agreements between the device vendors and solution provider are established.  When a customer wishes to purchase a solution, they generate a request for resource(s). The Solution Provider executes corresponding SLA and the service (if available) is allocated to the customer.  As soon as the SLAs are established, that is, appropriate smart contracts are executed, the service is started. The details of the resource provisioning are   out-of-scope of this work and interested readers can refer to~\cite{faisal2023accountable} for details.

Our goal is to ensure that the allocated resources (e.g., devices) perform as promised in their SLAs throughout their lifecycle. They must provide accurate data for  management purposes, such as billing and charging. To achieve this objective, it is imperative that the  devices must not be tempered by any party in any manner. To ensure this, we implement the following steps: \begin{figure}[!ht]
     \centering
   \includegraphics[width=70mm]{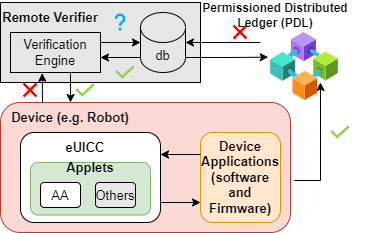}
    \caption{DIMSIM -- Modular Architecture of DIMSIM consists of 1) an Attestation Applet, 2) a Remote Verifier and c) a Permissioned Distributed Ledger}
    \label{fig:arch}
\end{figure}

\subsection{Initial Benchmark Measurements}

First step is to get the record initial measurements of the device software/firmware. To obtain the initial measurements, the hash value(s) of the software and firmware are  recorded in the PDL in the form of smart contract. Note that, the hash value is the complete hash of the software or firmware code. We advocate for a trustable and auditable system; therefore, the software and firmware installed on the device must be available on a location accessible by all stakeholders (e.g., edge cloud, git repository) along with their respective hash values. To protect Intellectual Property of the software, only the hash value of the software/firmware code may be available.

Once the solution is delivered to a customer's premises, the solution provider provisions the devices (e.g., robots). Device provisioning includes provisioning an operator profile and verifying the device software and firmware. Additionally, during the device provisioning, a secure element is also installed and configured as the `Attestation Applet' (AA) -- our novel applet, which is designed to monitor device integrity. The Attestation Applet hashes the device firmware and software, send these measurements to the solution provider.

\begin{figure}[!ht]
   \centering
    \includegraphics[width=90mm]{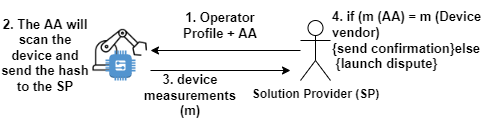}
    \caption{Attestation Applet provisioning and Initial Measurements Process}
    \label{fig:init_measurements}
\end{figure}

To confirm the device software and firmware are in the same state as the device vendor has promised in the contract, the solution provider compares the measurements with those provided by the device vendor with the measurements submitted by the Attestation Applet. If the measurements match, the solution provider will send a confirmation receipt to the AA, and the measurements sent by the AA, will be recorded as initial measurements inside the applet's non-volatile memory. The AA will use these values as the benchmark values for later comparison. The process is depicted in  Fig(\ref{fig:init_measurements}).

Since the SLA, that is, the smart contract recorded in the PDL, contains the agreed-upon measurements, alternatively, the AA can send measurements directly to the PDL. However, in such a case, AA will have to wait for transaction approval, which can take longer. Additionally,  PDLs are limited by the bandwidth they can handle, when many devices send requests, it can lead to congestion and increased risk of transaction rejection.

\subsection{Periodic Measurements with Attestation Applet}

After the initial measurements are recorded inside the device and the solution is initiated, the Attestation Applet scans the device firmware and software periodically and matches with the values recorded in the non-volatile memory from the previous step~(Fig.~\ref{fig:arch}). 

If the values measured by the AA  match with the recorded values, AA updates its $log(t\_s,current\_hash, previous\_hash, action\_taken$) with previous hash value similar to the current hash and action taken as $null$~(Table~\ref{tab:action_commands}) and waits for the next epoch to scan. If the values do not match, the AA updates its log and sends a message to the Remote Verifier. Optionally, if programmed, the AA can take preventive measures on its own such as stopping the device and terminate connectivity.

When the dispute message is received from the Attestation Applet, the verifier checks its own database for the records. If the `disputed hash' matches with its records, the remote verifier notifies the AA to update its records. If they do not, the remote verifier can take further actions such as stopping the device through control messages to the eUICC. 

\textbf{Note: Dispute Data Packet Details}
Dispute data packet sent by the Attestation Applet consists of the following fields: 
1) \textbf{$DeviceID~(5-7 bytes)$} -- device identifier, 2) \textbf{$AppletID~(5-7 bytes)$} -- applet identifier, 3) \textbf{$TimeStamp~(7 - 13 bytes)$}, 4) \textbf{$CurrentHash~(32 bytes)$} --  hash value calculated by the AA and  did not match with its stored and correct measurements, 5) \textbf{$PreviousHash~(32 bytes)$} the hash value  in the previous epoch, 6) \textbf{$ActionTaken~(32 bytes)$} is the device's immediate action after the scan result. The possible values for Action Taken are listed in Table~\ref{tab:action_commands}.

\begin{table}[!h]
    \centering
    \begin{tabular}{|l|l|l}
    \cline{1-2}
    \multicolumn{1}{|c|}{\textbf{ID}} & \multicolumn{1}{c|}{\textbf{Action}}       &  \\ \cline{1-2}
    0x00                              & \textit{null}              &  \\ \cline{1-2}
    0x01                              & \textit{Initiate investigation}                     &  \\ \cline{1-2}
    0x02                              & \textit{Restrict application or software execution} &  \\ \cline{1-2}
    0x03                              & \textit{Isolate device}                            &  \\ \cline{1-2}
    0x04                              &  \textit{Contain device}                             &  \\ \cline{1-2}
    0x05                              & \textit{Revoke device}                             &  \\ \cline{1-2}
   0x06                              & \textit{Stop and quarantine a file}                &  \\ \cline{1-2}
    0x07                              & \textit{Request deeper investigation}               &  \\ \cline{1-2}
    \end{tabular}
    \vspace{0.1cm}
    \caption{Commands for Action Taken }
    \label{tab:action_commands}
    \end{table}

    \subsection{Device Software/Firmware Updates}

    The solution provider setup service level agreement with device vendors. These SLAs executions are recorded in the PDL. 
    
    As devices are part of a solution, when a device vendor wants to update their software and/or firmware, the updates to one device can impact the performance of the complete solution, therefore solution provider must agree to the new software updates. To that end, device vendor notifies the solution provider with intended software/firmware updates and its hash value. If the  solution provider agrees with the updates, they will send a confirmation. The device vendor then executes the software update smart contracts with the updated software/firmware hash and confirmation receipt from the solution provider. The device vendor also sends a notification to remote verifier that the software/firmware is updated and send updated hash of the software/firmware to update their records. If the solution provider does not agree with the update, they can launch a dispute and settle as per the SLA. 

    \begin{figure}[!ht]
        \centering
        \includegraphics[width=80mm]{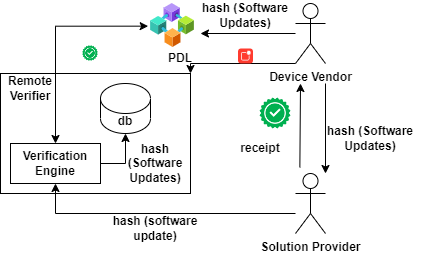}
        \caption{Software Updates on the Remote Verifier}
        \label{fig:software_update}
    \end{figure}

\section{Considerations}
\subsection{Storage}
Traditional Secure Elements  typically can store 4kb of application data~\cite{schlapfer2019security}. This means that, with the given capacity, our log file can hold up to 125 hashes, each consisting of 32 bytes, at any given time. Once this limit is reached, the log file will have to be archived a remote location, for example, transferred to the Remote Verifier's immutable database. 

EUICC depends on manufacturers design. New eUICC designs are still in the planning phase and the manufacturers are defining their proprietary applet sizes~\cite{eUICC_infinon}. As per the GSMA, the memory size of an eUICC can range from several kilobytes to several megabytes, with no specific limit on the number of applets that can be installed~\cite{gsma_euicc}. For the Attestation Applet, the storage requirement depends on a use case and the required frequency of device integrity verification is required.

\subsection{Response Time}

In DIMSIM, the AA will notify the remote verifier for any unexpected measurements. However, once the AA identifies a corrupted file and then notifies the remote verifier, in the meantime, a compromised/corrupted device can perform malicious activities, such as spreading a virus to its nearby devices. 

To that end, AA can block the device independently and before notifying the remote verifier. However, in such a setting, the device will have to be stopped for false alarms. 

\subsection{Corrupted Remote Verifier}
In DIMSIM, the Attestation Applet, sends a disputed hash to the remote verifier for the verification. If the remote verifier is compromised, it can reject the claims from the AA and allow a corrupted device to continue its operations. 

Recall that, the AA has direct access and authority to stop the device. The AA can be programmed in such a way to take preventive measures itself. For example,  after multiple alarms of malfunctioning, the AA can stop the device and send a notification to the solution provider. In another solution example, multiple remote verifiers can be deployed, and action will be taken only after collecting consensus from all the remote verifiers. However, multiple remote verifiers may introduce additional delays to the system.

\section{Evaluation setup \& preliminary Results}
Our proposal has two main components: \textbf{1)} A device equipped with eUICC and  \textbf{2)} the Remote verifier.

\subsection{An IoT Device equipped with eUICC}
Our objective is to ensure the device integrity. We achieve this through secure applets installed inside an eUICC. The AA scans the device periodically and matches with its own records. 

Our primary experimental setup is based on a Raspberry pi Model 4B, 4G LTE base Hat, Quectel EC25-E 4g/LTE module and Comprion test eSIM/eUICC card. Provision of the eUICC is done by the Nokia iSIM secure connect platform. The eUICC is connected indirectly via a modem.  Specific AT (Attention)  commands are used to enable the communication between the eUICC and a device. Application Protocol Data Unit (APDU) messages have been defined to interact within the Attestation Applet following the standard ISO7816.
Usually, eUICC supports different connectivity options to the host device.  The indirect communication via a modem, based on ISO7816 (T=0 protocol). The direct communication is based on protocols such as I2C or SPI. 

\begin{figure}[!h]
    \centering
    \includegraphics[width=80mm]{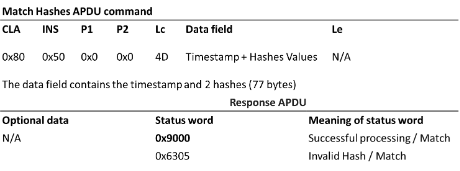}
   \caption{APDU Command to match hashes}
    \label{fig:results1}
\end{figure}

In Figure~\ref{fig:results1}, we present an example of hashes match command message used between the Attestation Applet and the device.

\subsection{Remote Verifier}
Our next objective is to assess the performance of the remote verifier and determine the cost associated with sending the hash value for verification. A crucial component of the remote verifier is its immutable database. Therefore, in our primary evaluations, we focus on both the hashing time and the expenses related to utilizing an immutable database.

The remote verifier has been implemented on an Ubuntu Virtual Machine with 4 vCPUs, 12 GB of RAM, and 149 GB of hard disk space. Its functionalities are coded in Python v3.10.2, utilizing an immutable database\footnote{https://immudb.io}. 

 We ran around 1000 iterations of our initial implementation and the analysis indicates that the 32 byte hash comparison at the remote verifier took maximum $\approx$203.941 $\mu$sec, $\approx$7.907 minimum, and the standard deviation is $\approx$9.875 $\mu$sec. We recorded data (Equation~\ref{eq1:data_format_for_db}) in the immutable database and the recording process  yields maximum value of $\approx$182.503 ms, minimum value is $\approx$8.447 ms with the standard deviation of $\approx$5.041 ms. Reading operations to the database take a lesser time with maximum value of $\approx$5.318 ms, minimum $\approx$0.583 ms.  and the standard deviation of $\approx$0.143 ms. As such, immudb is a lightweight and fast solution compared to other cryptographic-based data storage systems, such as blockchains~\cite{paik2020immudb}. While immudb utilizes cryptographic hashes to store records, which may result in lower performance compared to traditional databases, it's important to note that we primarily access the database for reading the latest values and resolving disputes. Data is only recorded in the event of software updates. 
\begin{equation}
    <software\_id, hashvalue, timestamp>
    \label{eq1:data_format_for_db}
\end{equation}


\begin{figure}[!h]
    \centering
    \includegraphics[width=100mm]{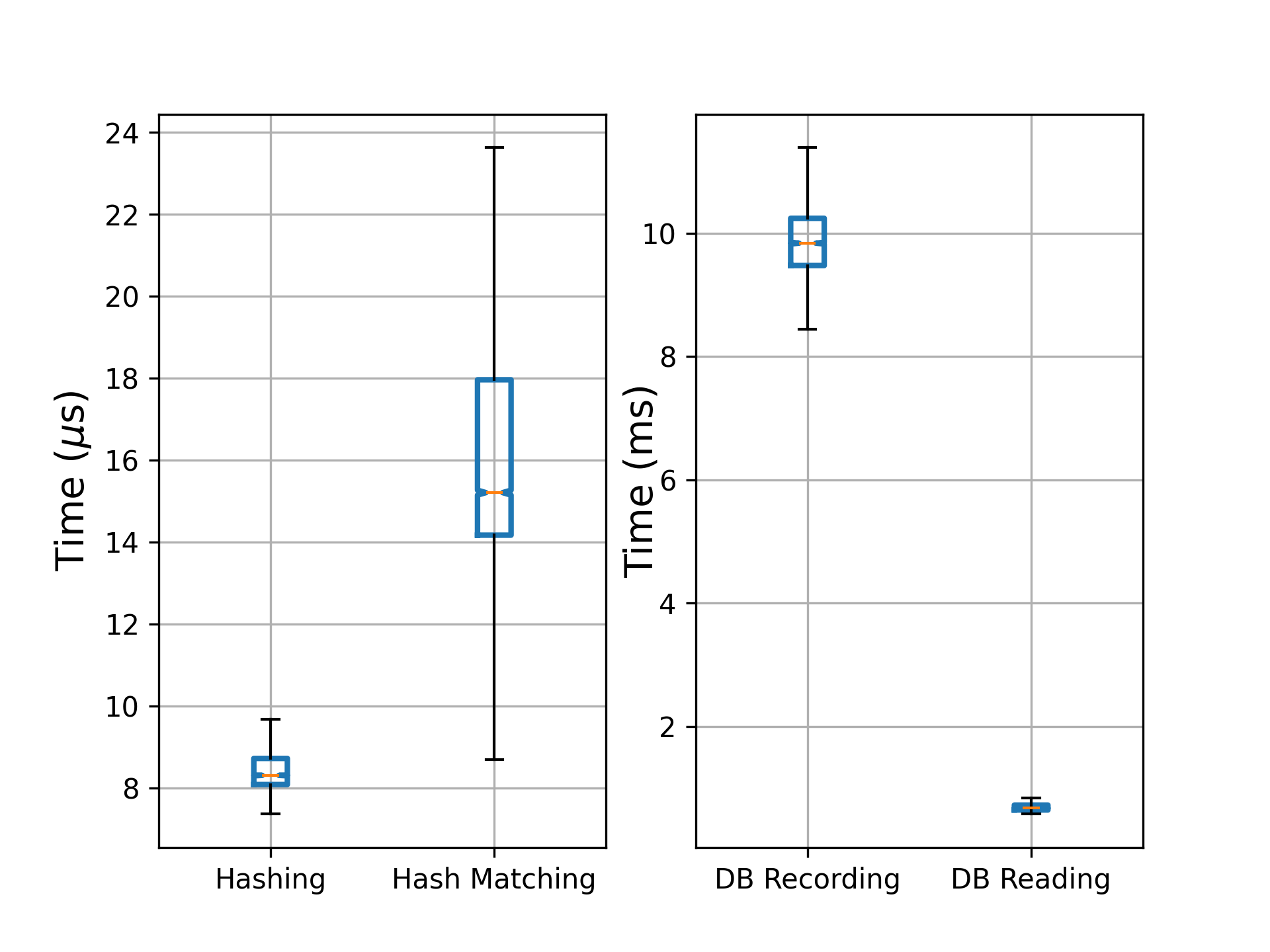}
    \caption{Remote Verifier Performance Evaluation - For hashing 32 bytes of data, the maximum value is $\approx$124.08 $\mu$sec, minimum value is $\approx$7.36 $\mu$sec and median is $\approx$8.304 $\mu$sec -- Immutable database evaluations show that recording data~(Equation~\ref{eq1:data_format_for_db}) took max $\approx$182.503 ms and minimum value is $\approx$8.447 ms and the median value  is $\approx$9.837 ms. The performance of write operation is a bit higher than the read operation, however, the write operation is used only in the situations of software updates}%
    \label{fig:rv_evaluations}%
\end{figure}

\section{Conclusion \& Future Work}

Ensuring the integrity of remote devices within  distributed ledger technology enabled systems  is of paramount importance. The challenge becomes particularly critical when it is employed for service level agreements, where the future settlements, such as, payments, rely on the data sent to distributed ledgers through smart contract executions. Currently, proposed methods require enormous amount of resources and thus are expensive to implement. In this paper, we present DIMSIM, our distributed architecture to ensure device integrity without the need for additional hardware, thereby facilitating inherently secure devices. 

We are currently in the process of developing a complete prototype for DIMSIM and plan to present our analysis in a future work. Importantly, when the future devices are integrated with our solution, their integrity will be guaranteed. Through our solution, the confidence and trust among the stakeholders will be enabled.

\bibliographystyle{IEEEtran}
\bibliography{bib/ref}
\begin{IEEEbiographynophoto}{Tooba Faisal} is a Research Web3 Engineer at Nokia Bell-Labs, France. She holds a PhD from King's College London along with an Master of Research in Security Science from University College London, an MS in Telecommunication and Networks and a BS in Computer Engineering from Bahria University. She has led ETSI ISG PDL report and specifications on smart contracts and currently serving as an expert in the ETSI Specialised Task Force (STF) on smart contracts. Her research interests encompass distributed ledger technology, Service Level Agreements,  Machine Learning and Next Generation Networks.
\end{IEEEbiographynophoto}
    
\begin{IEEEbiographynophoto} {Emmanuel Marilly}  is IoT Cloud Technical Leader at NOKIA Bell Labs Solution Research and is currently working on the Decentralized Operating System for Cyber Physical Systems project. He received a Ph.D. degree in computer science (Computer Vision, Neural Networks) at the University of Le Havre-Normandie, France. Prior to his tenure with NOKIA and Alcatel-Lucent, he was an assistant professor in computer science. For the past 20 years, he has worked on Internet of Things, Cloud Computing, DevOps, Machine Learning and A.I., Computer Vision, SLA management and Network and Service Management. Emmanuel is author of several papers and patents
\end{IEEEbiographynophoto}
\end{document}